\documentclass[reprint,superscriptaddress, amsmath,amssymb, aps]{revtex4-2}

\usepackage{graphicx}
\usepackage{dcolumn}
\usepackage{bm}
\usepackage{booktabs}
\usepackage[version=4]{mhchem}          
\usepackage[caption = false]{subfig}

\begin{document}

\preprint{APS/123-QED}

\title{Structure Studies of $^{13}\text{Be}$ from the $^{12}$Be(d,p) reaction in inverse kinematics on a solid deuteron target}
\author{J. Kovoor}
 
\author{K.L. Jones}%
\email{kgrzywac@utk.edu}
\author{J. Hooker}
\author{M. Vostinar}
\affiliation{Department of Physics and Astronomy, University of Tennessee, Knoxville, Tennessee 37996, USA}%
\author{R. Kanungo}
\affiliation{Astronomy and Physics Department, Saint Mary’s University, Halifax, Nova Scotia B3H 3C3, Canada}
\affiliation{TRIUMF, Vancouver, British Columbia V6T 4A3, Canada}
\author{S.D. Pain}
\affiliation{Physics Division, Oak Ridge National Laboratory, Oak Ridge, Tennessee 37831, USA}
\author{M. Alcorta}
\affiliation{TRIUMF, Vancouver, British Columbia V6T 4A3, Canada}
\author{J. Allen}
\affiliation{Physics Department, University of Notre Dame, Notre Dame, Indiana 46556, USA}
\author{C. Andreoiu}
\affiliation{Department of Chemistry, Simon Fraser University, Burnaby, British Columbia V5A 1S6, Canada}
\author{L. Atar}
\affiliation{Department of Physics, University of Guelph, Guelph, Ontario N1G 2W1, Canada}
\author{D.W. Bardayan}
\affiliation{Physics Department, University of Notre Dame, Notre Dame, Indiana 46556, USA}

\author{S.S. Bhattacharjee}
\affiliation{TRIUMF, Vancouver, British Columbia V6T 4A3, Canada}

\author{D. Blankstein}
\affiliation{Physics Department, University of Notre Dame, Notre Dame, Indiana 46556, USA}

\author{C. Burbadge}
\affiliation{Department of Physics, University of Guelph, Guelph, Ontario N1G 2W1, Canada}

\author{S. Burcher}
\affiliation{Department of Physics and Astronomy, University of Tennessee, Knoxville, Tennessee 37996, USA}

\author{W.N. Catford}
\affiliation{Department of Physics, University of Surrey, Guildford GU2 7XH, UK}

\author{S. Cha}
\affiliation{Department of Physics, Sungkyunkwan University, 2066 Seobu-ro, Jangan-gu, Suwon, Korea}

\author{K. Chae}
\affiliation{Department of Physics, Sungkyunkwan University, 2066 Seobu-ro, Jangan-gu, Suwon, Korea}

\author{D. Connolly}
\affiliation{TRIUMF, Vancouver, British Columbia V6T 4A3, Canada}

\author{B. Davids}
\affiliation{TRIUMF, Vancouver, British Columbia V6T 4A3, Canada}

\author{N. E. Esker}
\affiliation{TRIUMF, Vancouver, British Columbia V6T 4A3, Canada}

\author{F.H. Garcia}
\altaffiliation[Presently at ]{Lawrence Berkeley National Lab}
\affiliation{Department of Chemistry, Simon Fraser University, Burnaby, British Columbia V5A 1S6, Canada}

\author{S. Gillespie}
\affiliation{TRIUMF, Vancouver, British Columbia V6T 4A3, Canada}

\author{R. Ghimire}
\affiliation{Department of Physics and Astronomy, University of Tennessee, Knoxville, Tennessee 37996, USA}

\author{A. Gula}
\affiliation{Physics Department, University of Notre Dame, Notre Dame, Indiana 46556, USA}

\author{G. Hackman}
\affiliation{TRIUMF, Vancouver, British Columbia V6T 4A3, Canada}

\author{S. Hallam}
\affiliation{Department of Physics, University of Surrey, Guildford GU2 7XH, UK}

\author{M. Hellmich}
\affiliation{Astronomy and Physics Department, Saint Mary’s University, Halifax, Nova Scotia B3H 3C3, Canada}

\author{J. Henderson}
\affiliation{TRIUMF, Vancouver, British Columbia V6T 4A3, Canada}

\author{M. Holl}
\affiliation{Astronomy and Physics Department, Saint Mary’s University, Halifax, Nova Scotia B3H 3C3, Canada}
\affiliation{TRIUMF, Vancouver, British Columbia V6T 4A3, Canada}

\author{P. Jassal}
\affiliation{Astronomy and Physics Department, Saint Mary’s University, Halifax, Nova Scotia B3H 3C3, Canada}

\author{S. King}
\affiliation{Physics Department, University of Notre Dame, Notre Dame, Indiana 46556, USA}

\author{T. Knight}
\affiliation{Astronomy and Physics Department, Saint Mary’s University, Halifax, Nova Scotia B3H 3C3, Canada}

\author{R. Kruecken}
\affiliation{TRIUMF, Vancouver, British Columbia V6T 4A3, Canada}

\author{A. Lepailleur}
\affiliation{Department of Physics and Astronomy, Rutgers University, Piscataway, New Jersey 08854, USA}

\author{J. Liang}
\affiliation{Department of Physics and Astronomy, McMaster University, 1280 Main St. W, Hamilton, Ontario L8S 4M1, Canada}

\author{L. Morrison}
\affiliation{Department of Physics, University of Surrey, Guildford GU2 7XH, UK}

\author{P.D. O’Malley}
\affiliation{Physics Department, University of Notre Dame, Notre Dame, Indiana 46556, USA}

\author{X. Pereira-Lopez}
\affiliation{Department of Physics and Astronomy, University of Tennessee, Knoxville, Tennessee 37996, USA}

\author{A. Psaltis}

\altaffiliation[Presently at ]{Institut f\"ur Kernphysik, Technische Universit\"at Darmstadt, Darmstadt 64289, Germany}
\affiliation{Department of Physics and Astronomy, McMaster University, 1280 Main St. W, Hamilton, Ontario L8S 4M1, Canada}

\author{A. Radich}
\affiliation{Department of Physics, University of Guelph, Guelph, Ontario N1G 2W1, Canada}

\author{J. Refsgaard}
\affiliation{Astronomy and Physics Department, Saint Mary’s University, Halifax, Nova Scotia B3H 3C3, Canada}
\affiliation{TRIUMF, Vancouver, British Columbia V6T 4A3, Canada}

\author{A.C. Shotter}
\affiliation{School of Physics and Astronomy, University of Edinburgh, JCMB, Mayfield Road, Edinburgh EH9 3JZ, UK}

\author{M. Williams}
\affiliation{TRIUMF, Vancouver, British Columbia V6T 4A3, Canada}

\author{O. Workman}
\affiliation{Astronomy and Physics Department, Saint Mary’s University, Halifax, Nova Scotia B3H 3C3, Canada}
\affiliation{TRIUMF, Vancouver, British Columbia V6T 4A3, Canada}


\date{\today}

\begin{abstract}
The low-lying structure of $^{13}$Be has remained an enigma for decades. Despite numerous experimental and theoretical studies, large inconsistencies remain. Being both unbound, and one neutron away from $^{14}$Be, the heaviest bound beryllium nucleus, $^{13}$Be is difficult to study through simple reactions with weak radioactive ion beams or more complex reactions with stable-ion beams. Here, we present the results of a study using the $^{12}$Be(d,p)$^{13}$Be reaction in inverse kinematics using a 9.5~MeV per nucleon $^{12}$Be beam from the ISAC-II facility. The solid deuteron target of IRIS was used to achieve an increased areal thickness compared to conventional deuterated polyethylene targets. The Q-value spectrum below -4.4~MeV was analyzed using a Bayesian method with GEANT4 simulations. A three-point angular distribution with the same Q-value gate was fit with a mixture of $s$- and $p$-wave, $s$- and $d$-wave, or pure $p$-wave transfer. The Q-value spectrum was also compared with GEANT simulations obtained using the energies and widths of states reported in four previous works. It was found that our results are incompatible with works that revealed a wide $5/2^+$ resonance but shows better agreement with ones that reported a narrower width.
\end{abstract}

\maketitle


\section{\label{sec:level1}INTRODUCTION}

The neutron-rich beryllium isotopes provide a case study for studying the effects of weak binding and the continuum on the structure of the nucleus. Alpha clustering in \ce{^8Be}, leads to molecular structures when the neutrons are added to produce \ce{^9Be} and \ce{^10Be} \cite{alphaCluster}, and halo structures close to the dripline, notably the one-neutron halo in \ce{^11Be} \cite{TanihataBe11Halo}, and two-neutron halo in \ce{^14Be} \cite{TARUTINA200453,Descouvement14Be}.
The one-neutron subsystem of \ce{^14Be}, \ce{^13Be}, is unbound, making it a Borromean nucleus. The continuum structures of this unbound subsystem are essential to the understanding of the two-neutron halo nucleus. \\

As the neutron dripline is approached, it is usual to find an unbound odd-N nucleus followed by a bound even-N nucleus, owing to the pairing force.  In the case of  \ce{^13Be}, the shell gap breakdown at N=8 for neutron-rich nuclei might also contribute to the instability of \ce{^13Be}. This breakdown in N=8 results in the ground state of \ce{^12Be} being comprised of 70\% $sd$-shell intruder configuration with a $p$-shell configuration dominating the $0_2^+$ state \cite{Pain12Be}. Experiments by Kanungo, Chen, and Johansen \cite{KANUNGO12Be,CHEN2018412,Johansen12Be} have also reported the presence of a weak $s$-wave in the $0_1^+$ and $0_2^+$ states of \ce{^12Be}. The resulting low-lying $1/2^+$, $1/2^-$, and $5/2^+$ states in $^{13}$Be have been reported at different energies and orders in experimental studies, as summarized in Table \ref{prevExpTable}. Similarly, theoretical studies are not in agreement over the low-lying structure of this unbound nucleus.

In an experiment published in 1973 \cite{Bowman_13Beproof}, Bowman confirmed that \ce{^13Be} is unbound, as indicated in a 1966 isotope search \cite{Poskanzer_13Beindication}. Since then different experimental techniques have been used to study \ce{^13Be} including fragmentation reactions \cite{Thoennessen,CHRISTIAN2008101}, neutron knockout from \ce{^14Be} \cite{SIMON2007267,KONDO2010245,aksyutina}, proton knockout from \ce{^14B} \cite{Lecouey2004}, breakup \cite{randisi}, and transfer reactions \cite{KORSHENINNIKOV199553,CORSI2019134843}.
The most consistent finding, independent of the reaction mechanism, is a resonance at around 2~-~2.4 MeV above the \ce{^12Be}$+n$ threshold. There have been reports of a virtual $s$-wave state or resonance close to the threshold \cite{SIMON2007267,KONDO2010245, aksyutina,randisi}.  The location of the expected $p$-wave strength is more uncertain.

\begin{table*}
\caption{\label{prevExpTable} Previous studies of the low-lying structure of \ce{^13Be}, up to around 2 MeV. A strength is defined as a single resonance, or a virtual state, or a mixture of resonances, or a virtual state and a resonance. Strengths 1a and 1b are those closest to threshold and cannot be resolved in this work. Strength 2 is the well-known $d$-wave resonance at around 2~MeV. Where the literature reports scattering length instead of energy this is quoted in parentheses.}
\begin{ruledtabular}
\begin{tabular}{lllccc}
 \multicolumn{1}{c}{Author (year)}&\multicolumn{1}{c}{Reaction}& Beam energy  & \multicolumn{3}{c}{Energy above the threshold (MeV) or (a$_s$) and J$^{\pi}$}\\
 &&($AMeV$)&Strength 1a&Strength 1b
&Strength 2\\ 
\midrule
 Aleksandrov \textit{et al.} (1983) \cite{Aleksandrov1983}&$^{14}\text{C}+^{7}\text{Li}$ & 5.86 &-&-&$1.8$ \\
 Ostrowski \textit{et al.} (1992) \cite{Ostrowski1992}&$^{13}\text{C}+^{14}$C
 &24.09 &-&-&$2.01 (\frac{5}{2}^+\text{ or }\frac{1}{2}^-)$\\
 Korsheninnikov \textit{et al.} (1995) \cite{KORSHENINNIKOV199553}&$^{12}$Be+$^2$H
 &55&-&-&$2.0$\\
 Von Oertzen \textit{et al.} (1995) \cite{VONOERTZEN1995c129}&$^{13}\text{C}+^{14}$C&24&-&-&$2.01 (\frac{5}{2}^+)$\\
 Belozyorov \textit{et al.} (1998) \cite{BELOZYOROV1998419}&$^{14}$C+$^{11}$B&17.27&-&0.80&$2.02$\\
Thoenessen \textit{et al.} (2000) \cite{Thoennessen}&$^{9}$Be$+^{18}$O&80&$0.20(\frac{1}{2}^+)$ &$0.80(\frac{1}{2})$&$2.02(\frac{5}{2}^+)$ \\
Lecouey \textit{et al.} (2004) \cite{Lecouey2004}&$^{14}$B+C&41&-&$0.7(\frac{1}{2}^+)$&$2.4(\frac{5}{2}^+)$\\
Simon \textit{et al.} (2007) \cite{SIMON2007267}&$^{14}$Be+C&287 &(-3.2~fm$^{-1}$) $(\frac{1}{2}^+)$&-&$2.00(\frac{5}{2}^+)$ \\
 Kondo \textit{et al.} (2010) \cite{KONDO2010245}&$^{14}$Be+p& 69 & (-3.4~fm$^{-1}$) $(\frac{1}{2}^+)$&$0.51(\frac{1}{2}^-)$&$2.39(\frac{5}{2}^+)$\\
 Aksyutina \textit{et al.} (2013) \cite{aksyutina}&$^{14}$Be+p&304&$0.46(\frac{1}{2}^+)$&-&$1.95(\frac{5}{2}^+)$\\
 Randisi \textit{et al.} (2014) \cite{randisi}&$^{14,15}$B+$^{\text{nat}}$C&35&$0.40(\frac{1}{2}^+)$&$0.85(\frac{5}{2}^+)$&$2.35(\frac{5}{2}^+)$\\
 Marks \textit{et al.} (2015) \cite{marks}&$^{13}$B+$^{9}$Be&71&-&$0.73(\frac{1}{2}^+)$&$2.56(\frac{5}{2}^+)$\\
 Ribeiro \textit{et al.} (2018) \cite{ribeiro}&$^{14}$B+CH$_2$&400&$0.44(\frac{1}{2}^-)$&$0.86(\frac{1}{2}^+)$&$2.11(\frac{5}{2}^+)$\\
 Corsi \textit{et al.} (2019) \cite{CORSI2019134843}&$^{14}$Be+p&265&(-9.2~fm$^{-1}$) $(\frac{1}{2}^+)$&$0.48(\frac{1}{2}^-)$&$2.30(\frac{5}{2}^+)$\\
\end{tabular}
\end{ruledtabular}
\end{table*}


The first observation of a resonance in \ce{^13Be} was in 1983 \cite{Aleksandrov1983}, at 1.8~MeV above the neutron threshold. Multinucleon transfer studies \cite{Ostrowski1992} reported resonances at 2.01 and 3.12 MeV. Analysis of the transfer reaction using a \ce{^12Be} beam with a CD$_2$ target \cite{KORSHENINNIKOV199553} revealed resonances at approximately 2, 5, 7, and 10 MeV above the threshold.

Neutron knockout studies \cite{SIMON2007267,KONDO2010245,aksyutina,CORSI2019134843} using the invariant mass technique agree on a resonance around 2 MeV. After reconstructing the energy levels from the transverse momentum distribution of the \ce{^12Be}$+n$, Kondo \textit{et al.} \cite{KONDO2010245} reported a virtual $s$-wave state with a scattering length of -3.4 fm, a $p$-wave resonance at 0.51 MeV, and a broad resonance at 2.39 MeV. The low-lying $p$-wave can be interpreted as the quenching of the N=8 shell gap. Simon \textit{et al.} performed knockout from a $^{14}$Be beam on a carbon target and reported a $p$-wave resonance at 3.04 MeV, a $d$-wave resonance at 2.0 MeV, and a weak virtual $s$-wave state with a scattering length of -3.2 fm \cite{SIMON2007267}. Using data from \ce{^14Be} impinging on a liquid hydrogen target, combined with results from  earlier studies, Aksyutina \textit{et al.} reported an $s$-wave resonance at 0.45~MeV and $d$-wave resonance at 1.95~MeV, along with higher-lying states \cite{aksyutina}. Using a similar experimental method, Corsi \textit{et al.} \cite{CORSI2019134843} reported a virtual $s$-wave with a scattering length of -9.2 fm, as well as $p$-, and $d$- wave resonances at 0.48~MeV and 2.3~MeV, respectively.

A proton knockout study by Lecouey \textit{et al.}  revealed a structure around 0.7~MeV above the threshold, which was assigned as an $s$-wave, and a $d$-wave resonance at 2.4~MeV \cite{Lecouey2004}. Randisi \textit{et al.} used the coincident detection of neutrons and \ce{^12Be} from the breakup of \ce{^14B} and \ce{^15B} on a carbon target, and reconstructed the invariant mass of the \ce{^12Be}$+n$ and \ce{^12Be}$+n+n$ systems  \cite{randisi}. They have reported $s$-, $d$-, and $d$- wave resonances at 0.40~MeV, 0.85~MeV, and 2.35~MeV, respectively. Results of a proton knockout experiment from \ce{^14B} using a polyethylene target by Ribeiro \textit{et al.} \cite{ribeiro} was interpreted as revealing $p$-, $s$-, and $d$- wave resonances in \ce{13Be} at 0.44~MeV, 0.86~MeV, and 2.11~MeV, respectively.

Marks \textit{et al.} \cite{marks} performed a charge exchange reaction on \ce{^13B} and reported a low-lying $s$-wave resonance and a $d$-wave resonance at 2.56~MeV.

Various theoretical calculations including the shell model \cite{POPPELIER1985120,hTFortune13Be}, the Nilsson model \cite{macchiavelli}, and the relativistic mean field theory \cite{Ren1997_RElMFT}, have been used to study the \ce{^13Be} structure. Except for the agreement on a $5/2^+$ resonance, around 2~MeV above the neutron threshold, these studies do not agree on the position and parity of the low-lying states. There are large discrepancies in the width of the states between theoretically calculated and experimentally measured values. Using a microscopic cluster model, Descouvement \cite{DESCOUVEMONT_Theory} inferred a $1/2^+$ ground state, close to the threshold and a $d$-wave resonance slightly above 2~MeV. Antisymmetrized molecular dynamics calculations by Kanada-En'yo \cite{kanada_AMD} do not favor a $1/2^+$ ground state.

Here, we present the results from the first measurement of the \ce{^12Be}(d,p) reaction to study the low-lying states of \ce{^13Be}. The Q-value spectrum of \ce{^13Be} from the data was fit with a response function obtained from GEANT4 simulations using a Bayesian optimization procedure.

\section{EXPERIMENTAL SETUP}
The experiment was performed at the IRIS facility \cite{Kanungo2014} at TRIUMF, Canada. The radioactive $^{12}$Be ion beam, produced by the isotope separation online (ISOL) method after bombarding a tantalum target with a 500 MeV proton beam \cite{Dilling2014_ISAC}, was delivered by the ISAC-II superconducting linear accelerator. The secondary beam was accelerated to $9.5A$MeV with an average total intensity of 800~pps. The main components of IRIS are a low-pressure ionization chamber (IC) to provide initial beam particle identification to remove isobaric contaminants, a novel solid deuterium target, and charged particle detectors to detect the projectiles and recoils \cite{Kanungo2014}. The entire assembly was cooled to a temperature of 4K and deuterium gas was sprayed onto a 4.64 $\mu$m thick silver foil to give the desired thickness of deuterium, 52~$\mu$m for this experiment.


A 500 $\mu$m thick, annular YY1 (Yu) type (MICRON Semiconductors \cite{micronYY1}) silicon detector was placed 80.8 mm upstream of the target to detect the protons from the (d,p) reaction and covered a laboratory angular range of 122$^\circ$-148$^\circ$.  The 16 concentric rings were used to determine the scattering angle and the energy deposited in the strip was used to find the kinetic energy of the proton. The angles covered by the Yu detector correspond to forward angles in the center-of-mass, where low-lying levels with low orbital angular momentum transfer are expected to peak in cross section. Eight of these azimuthal detector sectors were joined together to form a complete annular ring. Low-energy, backward scattered protons were completely stopped in the detector, making it possible to reconstruct the reaction kinematics.

A similar YY1 detector (Yd) was mounted 86~mm downstream of the target to measure other reaction channels (d,d) and (d,t) that are mostly restricted to forward angles in inverse kinematics. The Yd silicon detector was backed by a CsI detector in a telescopic configuration for particle identification and to measure the full energy of higher-energy, forward scattered particles.
The scattered beam was detected in the S3-style annular silicon detectors \cite{micronS3} placed 600~mm and 690~mm  downstream of the target in a telescopic fashion to give particle identification of the recoils from the reaction.
All the silicon detector energies were calibrated using a triple alpha source (\ce{^239Pu}, \ce{^241Am}, and \ce{^244Cm}) giving $\alpha$-particle energies of 5.156~MeV, 5.486~MeV, and 5.805~MeV respectively.

A \ce{^12C} beam impinging on the solid deuterium target was used to benchmark the experimental setup since the levels in \ce{^13C} area are well-known; the results from the \ce{^12C} beam data are published in Hooker \textit{et al.} \cite{joshPaper}.  The \ce{^12C} and \ce{^12Be} beams were run for 3.6 and 4.2 days respectively. Runs without the target were used with both the \ce{^12C} and \ce{^12Be} beams to characterize the target and the background from the silver foil backing and other non-target sources.

\section{Results}
\begin{figure}[t]
\includegraphics[width=3.3in]{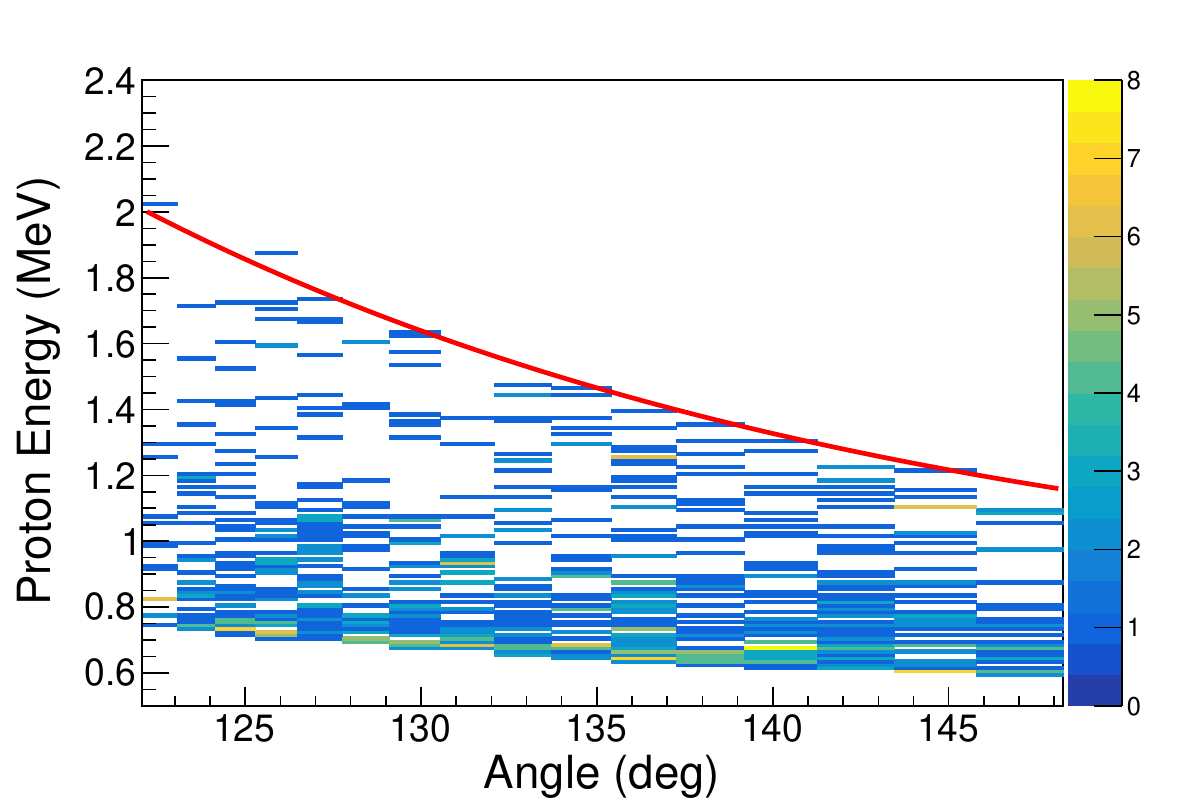}
\caption{\label{kinematics} Energy versus angle of protons in the laboratory frame for the \ce{^12Be}+d reaction, gated on a \ce{^12Be} ion detected in both the upstream ionization chamber and the downstream S3 telescope. The red curve shows the threshold for the ground state proton.}
\end{figure}
Figure \ref{kinematics} shows the energies of the protons measured in the upstream (Yu) detector, corrected for the energy loss at the center of the target, as a function of the angle. The data are gated on both a \ce{^12Be} beam particle, identified in the upstream ionization chamber, and a \ce{^12Be} residue identified in the downstream S3 telescope.
Loci from neutron transfer can be discerned in the kinematics plot shown in Figure \ref{kinematics}.  In particular, it is possible to trace a low-energy locus corresponding to the highest Q values populated in the reaction, and the spectrum is clean above this line. The clean area below 0.8~MeV at 122$^{\circ}$ tracing down to 0.6~MeV at 150$^{\circ}$ is due to the low-energy threshold imposed via software.

\begin{figure}[t]
\includegraphics[width=3.4in]{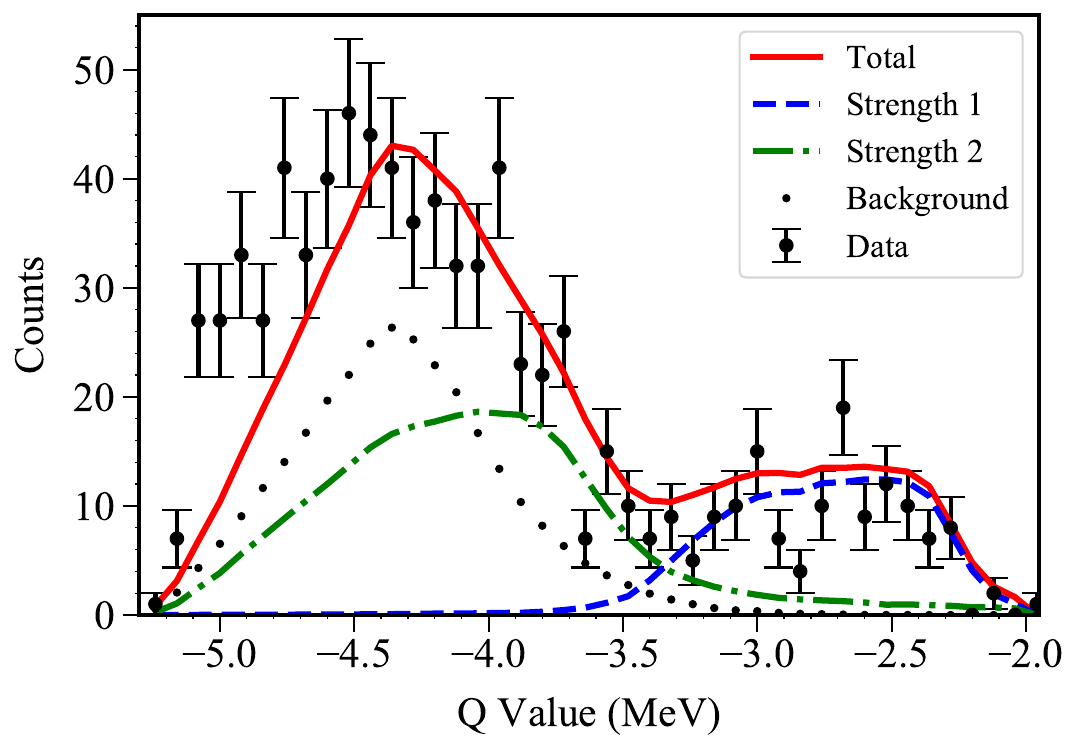}
\caption{\label{qValue} \ce{^13Be} Q-value spectrum obtained from the \ce{^12Be} + d reaction. The data is shown as black points with statistical error bars. A fit to the data obtained using Bayesian optimization is shown as a red line. Low-lying strength lying at 0.55~MeV above the threshold is denoted as a blue dashed line. Strength at 2.2~MeV above the threshold is shown as a green dot-dashed line. The non-resonant background is shown as a black dotted line.}
\end{figure}

The Q value obtained on an event-by-event basis is shown in Figure \ref{qValue} as data points with statistical error bars. There is a broad structure between $Q~=-3.2$~MeV and $-2.2$~MeV, which could also be interpreted as two peaks, and a clear increase in intensity below $Q=-3.4$~MeV.  To interpret the data further, it was necessary to use a GEANT4 simulation that accounted for the energy loss of the beam in the ionization chamber, the energy loss of the beam and the reaction protons in the solid deuterium target, the resolution of the Yu detector, and a non-resonant continuum.  After analyzing the data without the solid deuterium target, it was found that there was no background present in the angular region covered by the upstream YY1 detector resulting from the silver foil backing.  Therefore, this did not need to be included in the simulation.

The energy loss in the IC was simulated using the known thicknesses of the windows and the pressure of the isobutane gas which was 19.5 Torr. The thickness of the solid deuteron target was found by comparing the energy of the beam as measured in the downstream S3 telescope with the solid deuterium target in and out. The analysis procedure assumes that the reaction happened at the center of the target. The energy loss of the protons exiting the target was calculated and added back to the measured energy. The detector resolution was characterized using data from the alpha calibrations.  The low-lying resonances of unbound nuclei will sit on the background from the non-resonant continuum. This background was modeled using the fifth power of the energy above the neutron threshold, to reflect the increase in phase space. Each resonance was parameterized in the GEANT4 simulation as a Breit-Wigner line shape.
For consistency, GEANT4 energy loss tables were used to calculate the energy losses of the \ce{^12Be} beam, protons, and other charged particles throughout the experimental data analysis and the simulations. 

In this study, we used Bayesian optimization in combination with GEANT4 simulations to better understand the Q-value spectrum of \ce{^13Be}.
This spectrum was fitted using GEANT4 simulations described above, and a Bayesian optimization procedure as described and demonstrated for the \ce{^12C} beam data in Hooker \textit{et al.} \cite{joshPaper}. Bayesian Optimization (BayesOpt) is a tool to find the global extrema of arbitrary functions especially when it is computationally very expensive to evaluate these functions.

In this procedure, the positions, widths, and the intensity of the states along with the amplitude of the non-resonant background were given as free parameters. The best fit was defined as that which minimized the $\chi^2$ value between the measured and simulated spectra. Local minima were encountered when all parameters were allowed to vary without any bounds. Therefore, a random search was used, and the energies and the amplitudes of the states were found to produce the most sensitivity. Bounds for these parameters were found by using an iterative search and $\chi^2$ minimization. 

Two strengths were used in the fit, as shown in Figure \ref{qValue}. The resolution of the data was dominated by the energy loss and straggling of the emergent protons. As the Q value gets more negative, we can see that the peak gets broader. This is because the protons with the lowest energies lose proportionally more of their energy exiting the target. At the limit of the most negative Q value, the protons deposit all their energy in the target and cannot be detected.  These combined effects explain the asymmetric, non-Lorentzian shape of the resonances.


\begin{table}[ht]
\caption{\label{qValTable}%
The Q values and the widths of the states were obtained from the fit. Energies are in MeV.
}
\begin{ruledtabular}
\begin{tabular}{cccc}
\textrm{Strength}&
\textrm{Energy above threshold}&
\multicolumn{1}{c}{\textrm{Q value}}&
\textrm{Width}\\
\colrule
\\
1 & $0.55^{+0.08}_{-0.07}$ & $-2.75^{+0.08}_{-0.07}$ & $0.11^{+0.04}_{-0.05}$\\
\\
2 & $2.22^{+0.04}_{-0.05}$ & $-4.42^{+0.04}_{-0.05}$ & $0.40^{+0.03}_{-0.04}$\\

\end{tabular}
\end{ruledtabular}
\end{table}
The data in Figure \ref{qValue} are shown as black circles with statistical error bars. The red line shows the total fit of the data. The black dotted line depicts the non-resonant background, the blue dashed line denotes the lowest-lying strength, and the green dot-dashed line shows the well-known resonance which is around 2 MeV above the neutron threshold. At very negative Q-values, below -4.4~MeV, the fit does not well reproduce the straggling effects of the lowest energy protons emerging from the target and the proximity of the proton energies to the detector thresholds. 

The error bars for the centroids in energy for each strength were found by varying the energy in small steps until a $\chi^2$ value of $\chi^2_{min}+1$ was found for both the upper and lower bounds.  These were taken as the $1\sigma$ error bars. For the strength close to the threshold, the Q value was found to be -2.75~MeV with a width of $0.11$~MeV and the Q value of the higher-lying resonance was $-4.42$~MeV. The energies and widths with error bars are listed in Table \ref{qValTable}.

It was observed that when a third strength was added to the simulation, the fit did not improve and the intensity of this third strength was minimal. Using only one strength reproduces the $d$-wave resonance and misses the data points above -3.25~MeV in Q-value. The $\chi^2/N$ above -4.44~MeV in Q value including all the data in the fit and using only one strength is 3.5, compared with 2.9 for two strengths.




\begin{figure}[t]
\includegraphics[width=3.3in]{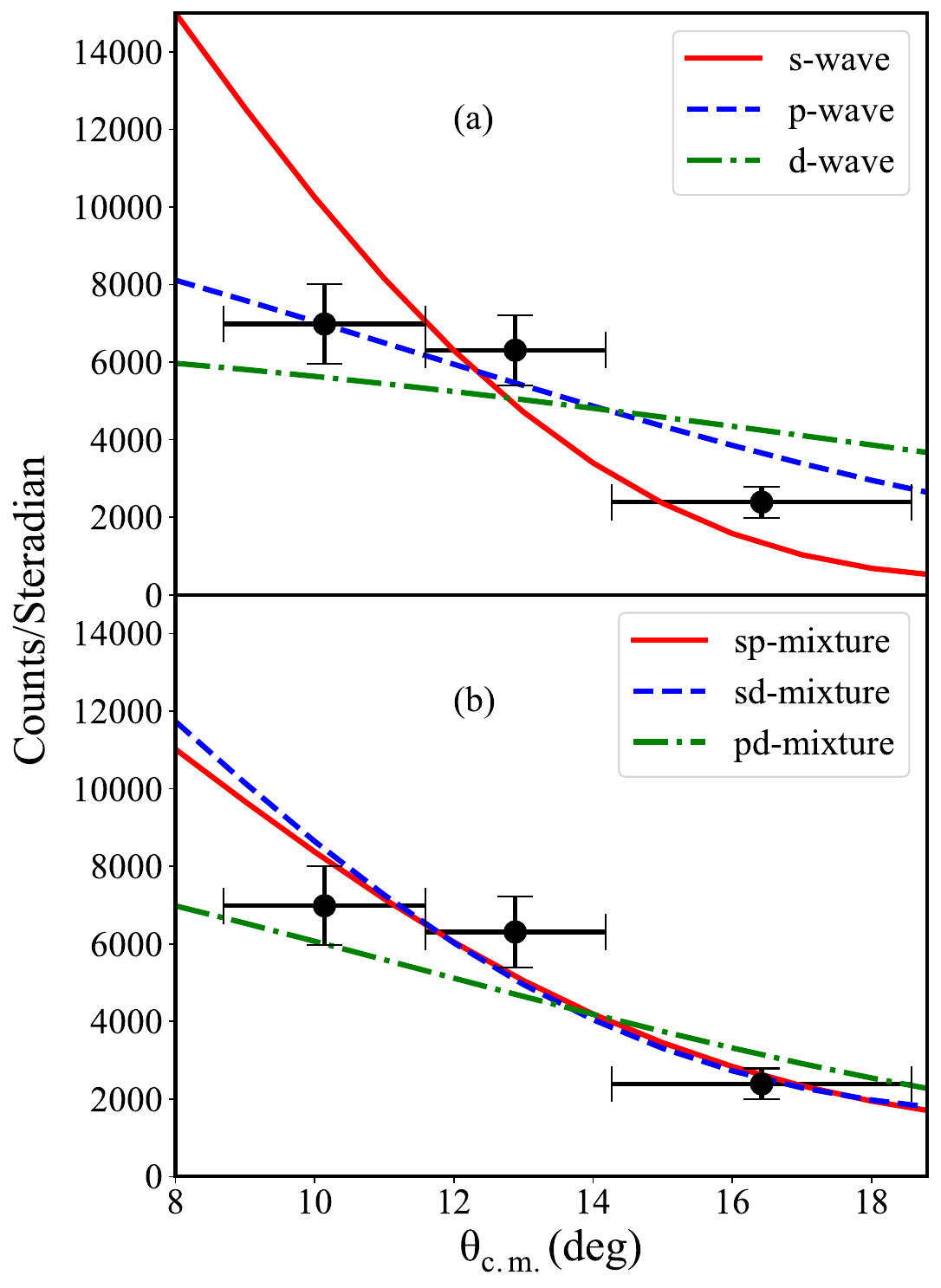}
\caption{\label{angDistWavesPPKD} Three-point angular distribution of protons in the center of mass frame in the Q-value range from -3.2 MeV to -2.2 MeV, for proton events with 0.55 MeV energy threshold.  The lines show DWBA fits to the data.  The calculations were performed using FRESCO using the Perey-Perey potential for the deuteron vertex and the Koning-Delaroche potential for the proton vertex.  Panel (a) shows data fitted assuming only one resonance is present and (b) assuming two unresolved resonances. The uncertainties in the counts and the angular range for the bins are denoted using the error bars.}
\end{figure} 

The angular distributions of the outgoing protons from the \ce{^12Be}(d,p) reaction give information about the spin and parity of the energy level of the transferred neutron. Owing to the low number of reaction protons measured, the angular distribution consists of three angular bins ranging from 121.8$^\circ$ to 131.8$^\circ$, 131.8$^\circ$ to 138.9$^\circ$, and 138.9$^\circ$ to 148$^\circ$, in the laboratory frame.

The reaction was analyzed in the distorted-wave Born approximation (DWBA) framework, using the FRESCO code \cite{THOMPSON1988167_fresco}. The optical potential parameters were obtained from FRONT \cite{front}. This method is sensitive to the specific optical potentials used. To account for this ambiguity, we used different optical potentials and studied the sensitivity of the fit to the data. 



Optical model calculations for the transfer reaction were calculated using Lohr Haeberli (LH) \cite{LOHR1974381} and Perey-Perey (PP) \cite{PEREY19761} potentials for the deuteron-nucleus vertex and for the proton-nucleus vertex, Koning and Delaroche \cite{KONING2003231} potential was used. The parameters used for the LH potential were radius, r~=~1.29~fm, diffuseness, a~=~0.860~fm, and a spin-orbit term, with $V_{SO}$~=~3.5~MeV and for PP,  r~=~1.29~fm, diffuseness, a~=~0.810~fm. The Perey-Perey potential does not include a spin-orbit term. For the lowest-lying strength (the blue dashed line in Figure \ref{qValue}), the angular distribution was evaluated by dividing the data into the three angular bins and the number of integrated counts was obtained for each bin. The non-resonant continuum was subtracted from the data. 
After computing the ratio between the counts and the solid angle coverage for each bin and converting it into the center-of-mass frame, the resulting distribution was compared to the FRESCO calculations. The angular distributions for the \ce{^12Be}(d,p)\ce{^13Be} reaction were fitted with pure $s$-, $p$-, and $d$-wave transfer as well as for two unresolved resonances with different amounts of transferred angular momentum, namely $s$- and $p$-wave, $s$- and $d$-wave, and $p$- and $d$-wave transfer mixtures. The fit was performed using Orthogonal Distance Regression (ODR) to account for the error in both the counts and the angular bins. The fits using the Perey-Perey and the Koning-Delaroche potentials are shown in Figure \ref{angDistWavesPPKD}.\\
\begin{table*}[ht]
\caption{\label{LH_KD_Table}%
$s$-, $p$-, and $d$-wave components of the lowest-lying strength (Strength 1) extracted from $\chi^2$ fitting of DWBA calculations, using Lohr-Haeberli and Perey-Perey deuteron optical potentials, to the experimental angular distributions. Koning-Delaroche proton optical potentials were used in all cases. Fractions 1 and 2 stand for the first and second wave components in the mixture.
}
\begin{tabular}{rccccccc}
\toprule
 & \multicolumn{3}{c}{\textrm {LH + KD}} & & 
\multicolumn{3}{c}{\textrm{PP + KD}} \\
\cmidrule(r){2-4} \cmidrule(l){6-8}
\textrm{Mixture} & 
\textrm{Fraction 1} & \textrm{Fraction 2} & $\chi^2$/N & &
\textrm{Fraction 1} & \textrm{Fraction 2}& $\chi^2$/N\\
\midrule
s  & 1.00 & N/A  & 5.66 && 1.00   & N/A  & 8.97\\
p  & 1.00 & N/A  & 5.66  && 1.00   & N/A  & 5.41\\
d  & 1.00 & N/A  & 13.64  && 1.00   & N/A  & 12.59\\
sp & $0.70^{+0.08}_{-0.06}$ & $0.30^{+0.06}_{-0.08}$ & 3.32  && $0.65^{+0.09}_{-0.06}$   & $0.35^{+0.06}_{-0.09}$ & 3.3\\
sd & $0.93^{+0.02}_{-0.02}$ & $0.07^{+0.02}_{-0.02}$ & 3.87  && $0.92^{+0.015}_{-0.025}$   & $0.08^{+0.025}_{-0.015}$ & 4.01\\
pd & $1.00_{-0.035}$ & $0.00^{+0.035}$ &  11.32 && $1.00_{-0.025}$  & $0.00^{+0.025}$ & 10.82\\
\bottomrule
\end{tabular}
\end{table*}
As shown in Table \ref{LH_KD_Table}, a mixture of $s$- and $p$-wave transfers provided the lowest $\chi^2/N$ value for both sets of calculations. Here, N is the degrees of freedom; $N=2$ for single wave fits and $N=1$ for fits using mixtures. With the Perey-Perey potential, the best fit gave $65\%^{+9\%}_{-6\%}$ $s$-wave and $35\%^{+6\%}_{-9\%}$ $p$-wave, with a $\chi^2/N$ of 3.3. For the Lohr-Haeberli potential, the best fit gave a combination of $70\%^{+8\%}_{-6\%}$ $s$-wave and $30\%^{+6\%}_{-8\%}$ $p$-wave and the $\chi^2/N$ was very similar to the PP case at 3.32.
The next best fit was with a mixture of $s$- and $d$-wave transfers. The Perey-Perey potential gave $92\%^{+1.5\%}_{-2.5\%}$ $s$-wave and $8\%^{+2.5\%}_{-1.5\%}$ $d$-wave, while the Lohr-Haeberli potential gave a mixture of $93\%^{+2\%}_{-2\%}$ $s$-wave and $7\%^{+2\%}_{-2\%}$ $d$-wave. The $\chi^2/N$ values were 4.01 and 3.87 for the LH and PP potentials respectively. The $1\sigma$ error bars were calculated by varying the fit parameters which produced a $\chi^2$ value of $\chi^2_{min}+1$.

The fits assuming a single virtual state or resonance gave somewhat higher $\chi^2/N$ values compared to the assumption of two unresolved virtual states or resonances. The pure $p$-wave transfer fitted the data better than either pure $s$-wave, or $d$-wave transfer with the PP potential, but with the LH potential, both $p$-, and $s$-waves produced identical $\chi^2$ values.


The interpretations of the low-lying structure of $^{13}$Be from four recent experiments were compared with the Q-value spectrum from the current data. The four experiments chosen for the comparison were Kondo \textit{et al.} \cite{KONDO2010245}, Corsi \textit{et al.} \cite{CORSI2019134843}, Ribeiro \textit{et al.} \cite{ribeiro}, and Randisi \textit{et al.} \cite{randisi}.  The first two used neutron knockout reactions with 69~AMeV and 265~AMeV  \ce{^14Be} beams, respectively, impinging on cryogenic hydrogen targets at RIKEN. Ribeiro performed a proton knockout reaction from a 400~AMeV \ce{^14B} beam at GSI, and Randisi \textit{et al.} used the same beam-target combination, but at a lower beam energy, 35~AMeV at GANIL.  The Randisi experiment also used the breakup of a \ce{^{15}B} beam. 
\begin{figure*}[t]
{\includegraphics[width=6.in]{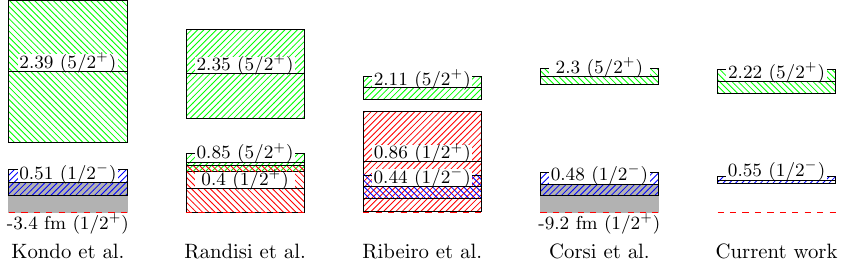}}
\caption{Schematics showing the comparison between the results of previous works and the current work. In the current work, only the case with a pure $p$-wave is shown as the position of the waves in the mixtures could not be resolved. Comparisons are made with Kondo \textit{et al.} and Corsi \textit{et al.} who both reported a virtual $s$-wave state, and $p$- and $d$-wave resonances, and with Randisi \textit{et al.} and Ribeiro \textit{et al.} showing $s$-, $p$-, and $d$-wave resonances. Red, blue, and green lines depict $s$-, $p$-, and $d$- waves respectively. The red-dashed lines show the threshold. The gray shaded region shows the presence of a virtual state.}
\label{someexample}
\end{figure*}
The reactions with a $^{14}$Be beam and a proton target provide a clean, simple reaction mechanism, but can suffer from the issue of identifying the neutron associated with $^{13}$Be from the one removed in the reaction in cases where both neutrons exit the target. Both the experiments performed by Kondo \textit{et al.} \cite{KONDO2010245} and Corsi \textit{et al.} \cite{CORSI2019134843} included $\gamma$-ray detection and the interpretations include the effects of an excited $^{12}$Be component to the ground-state wave function of $^{14}$Be.  Proton removal in $^{14}$B allows the unambiguous identification of the neutron from $^{13}$Be. However, the reaction mechanism is less well described by the sudden approximation owing to the extra binding of the last proton in $^{14}$B.


Figure \ref{someexample} illustrates interpretations from the four previous measurements compared with that from the fit to the current data assuming strength 1 comes from a single $p$-wave resonance. The lower $\chi^2$ fits with two states in strength 1 cannot uniquely locate the centroids of the two states and are therefore not useful in this comparison. The position and width of the $d$-wave resonance at 2.22~MeV from the current analysis is similar to that found by both \citet{CORSI2019134843} and \citet{ribeiro}. \citet{KONDO2010245} and \citet{randisi} found a larger width for this resonance. Furthermore, \citet{KONDO2010245,CORSI2019134843,ribeiro} all place a $\frac{1}{2}^-$ resonance between 0.44 and 0.51~MeV, with $s$-wave strength closer to threshold, and \citet{randisi} interpreted the near-threshold structure as a narrow $d$-wave resonance and a broader $s$-wave level.

\begin{figure*}
\includegraphics[width=6.6in]{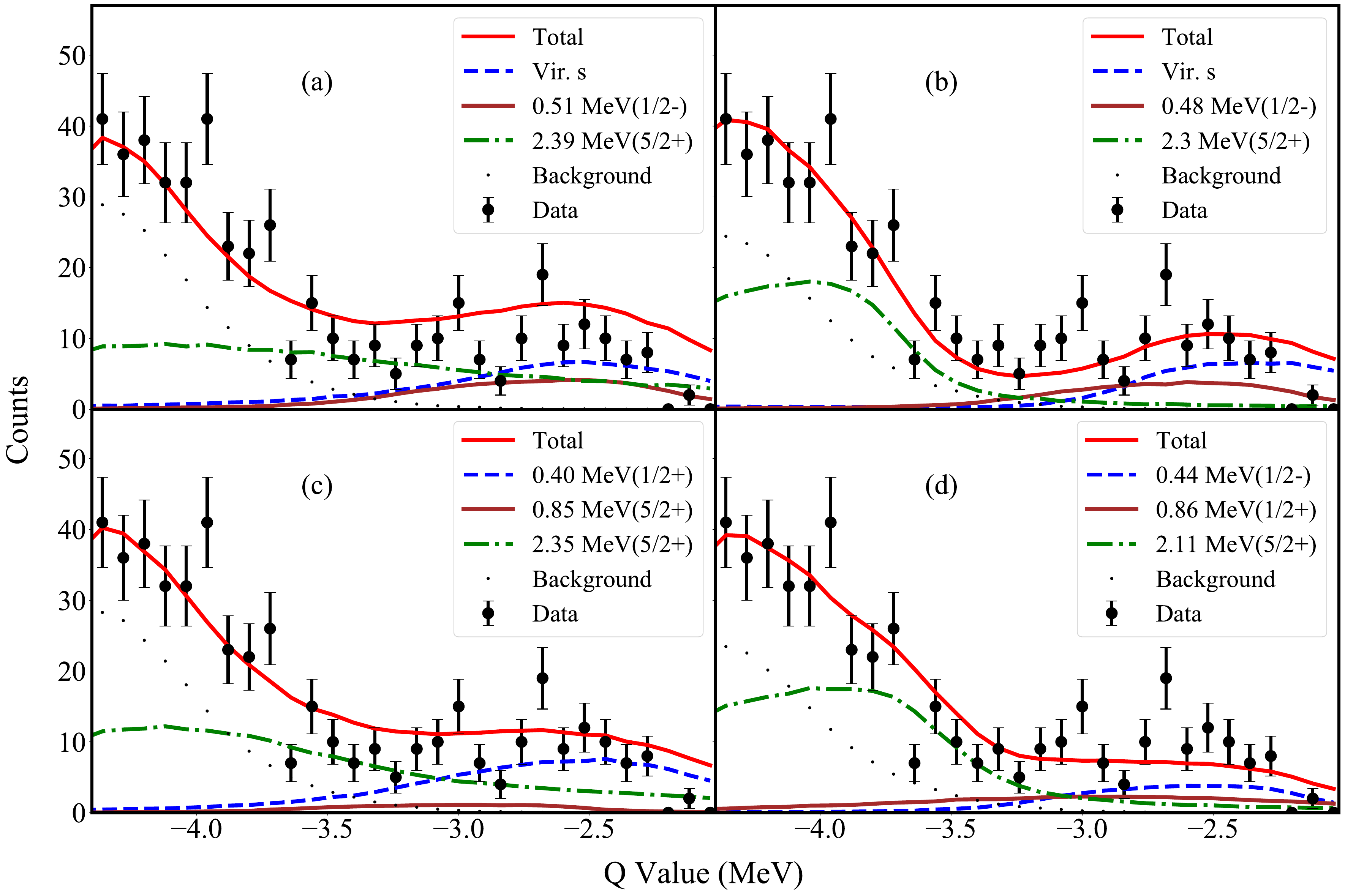}
\caption{\label{QvalMatrix} Data fitted with GEANT4 simulations with energy and widths obtained from (a) Kondo \textit{et al.} \cite{KONDO2010245}, (b) Corsi \textit{et al.} \cite{CORSI2019134843}, (c) Randisi \textit{et al.} \cite{randisi}, (d) Ribeiro \textit{et al.} \cite{ribeiro}. The amplitudes of the states were used from the angular distributions. The global fit is shown as the red line and the background is denoted as black dots. The lowest-lying strength is shown as blue-dashed line irrespective of its nature. The higher lying states are depicted as solid brown and green dot-dashed lines.}
\end{figure*}

To further these comparisons, our data were fit assuming strength 1 was an unresolved combination of two resonances, or of a resonance and a virtual state. This was necessary as all four of the previous measurements in the comparison are interpreted as having two resonances or a resonance and a virtual state below $\sim$ 2~MeV. These were named strengths 1a and 1b. The positions and widths for these states used in this set of simulations came from the four previous experiments discussed above. 
The relative amplitudes of the states and resonances were constrained in the GEANT4 simulations using the values in Table \ref{LH_KD_Table} from the angular distribution fit using Perey-Perey and Koning-Delaroche potentials. The simulated spectrum was then scaled to the data. Bayesian optimization was not carried out in this part of the analysis.  The results from the fits using the positions and widths of states from literature are shown in Figure \ref{QvalMatrix}. The fits using values from both Ribeiro \textit{et al.} and Corsi \textit{et al.} give similar $\chi^2/N$ values, 2.77 and 2.62, respectively, and they are the lowest of all the four. The widths reported by Kondo \textit{et al.} and Randisi \textit{et al.} for the $\frac{5}{2}^+$, $\sim2$~MeV resonance are 2.4~MeV and 1.5~MeV, respectively. They are too large to agree with our data thus giving higher $\chi^2/N$ values, 5.51 and 3.47, respectively.

\section{Discussion}


Neutron transfer on a $^{12}$Be beam provides a clean probe of the structure of $^{13}$Be. The energy loss of protons exiting the solid deuterium target limited the resolution in this measurement, necessitating the use of GEANT4 simulations as inputs into a Bayesian optimization to interpret the data.  Fitting a three-point angular distribution gated on the first 0.55~MeV above threshold in the Q-value spectrum revealed three possible interpretations of the low-lying strength: 1) a single $p$-wave resonance, 2) a strong $s$-wave state and a weaker $p$-wave resonance, or 3) a strong $s$-wave state with a weaker $d$-wave resonance.  In cases 2) and 3) the $s$-wave accounted for more than 60\% of the strength below 2~MeV.

Comparing the results from the current data in isolation with previous results showed agreement over the location of the higher-lying $d$-wave resonance. However, two (Refs. \cite{KONDO2010245} and \cite{randisi}) of the four previous measurements used in the comparison indicate a much broader $d$-wave structure than found in our work.

To make a complete comparison with previous results, the interpretations from the four measurements were used as inputs in the GEANT4 simulation of our experiment. The Q-value spectra from those simulations were compared with the measured Q-value spectrum. The best agreement was found for Corsi \textit{et al.} and Ribeiro \textit{et al.}, despite having different centroids and widths for the lowest $s$-wave state. The much broader $d$-wave resonances used to interpret the results from Kondo and Randisi are incompatible with our measurement.

This measurement, combined with those from Corsi \textit{et al.} and Ribeiro \textit{et al.}, strongly suggest that, in addition to the $d$-wave resonance at between 2.11 and 2.3~MeV, there is a $p$-wave resonance at between 0.44 and 0.55~MeV, and an additional less understood $s$-wave strength.  A future neutron-transfer measurement at a higher beam energy could provide the better resolution in Q value required to constrain the nature of this low-lying $s$-wave strength.

\begin{acknowledgments}
This research was supported by the U.S. Department of Energy, Office of Science, Office of Nuclear Physics under Contract No. DE-FG02-96ER40963 (UTK), DE-AC05-00OR22725 (ORNL), and the U. S. National
Science Foundation under Award Numbers PHY-1404218 (Rutgers)
and PHY-2011890 (Notre Dame). The authors are grateful for support from NSERC, Canada Foundation for Innovation and Nova Scotia Research and Innovation Trust, RCNP, grant-in-aid program of the Japanese government. TRIUMF is supported by a contribution through the National Research Council, Canada. The authors are thankful to the TRIUMF ISAC beam delivery team for providing the \ce{^12Be} beam. This work was supported by the National Research Foundation of Korea (NRF) grant funded by the Korea government
(MSIT) Nos. 2020R1A2C1005981 and 2016R1A5A1013277. This work
was partially supported by STFC Grant No. ST/L005743/1 (Surrey).
\end{acknowledgments}



\bibliography{13BePaper}
\end{document}